\documentclass[letterpaper, 10pt, conference]{ieeeconf}  

\usepackage{graphicx,keyval,trig,url} 
\usepackage[T1]{fontenc}     
\usepackage[utf8]{inputenc}	
\usepackage{graphicx}

\usepackage{amsthm}
\usepackage{amssymb}
\usepackage{amsmath}
\usepackage{mathtools}
\usepackage[mathscr]{eucal}
\usepackage{lipsum}
\usepackage{cuted} 
\usepackage{blindtext}
\usepackage{color}
\usepackage{dsfont}

\usepackage{epstopdf}
\usepackage{bm}   

\usepackage{color}
\usepackage{xcolor}

\usepackage{multirow}
\usepackage{longtable}

\usepackage{mathtools}

\IEEEoverridecommandlockouts                              

\title{\LARGE \bf
Periodic Load Rejection for Floating Offshore Wind Turbines via Constrained Subspace Predictive Repetitive Control*
}

\author{Yichao Liu$^{1}$, Riccardo M.G. Ferrari$^{1}$ and Jan-Willem van Wingerden$^{1}$
\thanks{*This research was supported by the European Union via a Marie Sklodowska-Curie Action (Project EDOWE, grant 835901).
$^{1}$Delft University of Technology, Delft Center for Systems and Control, Mekelweg 2, 2628 CD Delft, The Netherlands.
        {\tt\small \{Y.Liu-17, R.Ferrari, J.W.vanWingerden\}@tudelft.nl}.}}

\begin{document}

\maketitle
\thispagestyle{empty}
\pagestyle{empty}

\begin{abstract}
Individual Pitch Control (IPC) is an effective control strategy to mitigate the blade loads on large-scale wind turbines. 
Since IPC usually requires high pitch actuation, the safety constraints of the pitch actuator should be taken into account when designing the controller.
This paper introduces a constrained Subspace Predictive Repetitive Control (SPRC) approach, which considers the limitation of blade pitch angle and pitch rate.  
To fulfill this goal, a model predictive control scheme is implemented in the fully data-driven SPRC approach to incorporate the physical limitations of the pitch actuator in the control problem formulation.
An optimal control law subjected to constraints is then formulated so that future constraint violations are anticipated and prevented.
Case studies show that the developed constrained SPRC reduces the pitch activities necessary to mitigate the blade loads when experiencing wind turbulence and abrupt wind gusts.
More importantly, the approach allows the wind farm operator to design conservative bounds for the pitch actuator constraints that satisfies safety limitations, design specifications and physical restrictions.
This will help to alleviate the cyclic fatigue loads on the actuators, increase the structural reliability and extend the lifespan of the pitch control system.

\end{abstract}

\section{Introduction}
In past decades, wind energy has expanded by leaps and bounds in the international energy mix~\cite{Veers_2019}. 
The global wind industry reached a milestone of 651\,GW cumulative installed capacity in 2019, with a rapid growth of 10\% compared to 2018 \cite{GWEC_2020}. 
As onshore wind industry has matured, offshore wind is embraced by governments and picks up momentum in the race in achieving a low-carbon future \cite{Decastro-2019}. 

In particular, Floating Offshore Wind Turbines (FOWTs) offer a viable solution to tap higher wind speeds in deep waters and avoid the constraints imposed on land-based wind turbines \cite{Liu-2016}.
However, one of the main challenges in the development of FOWTs is the high Operation \& Maintenance (O\&M) cost \cite{Santos_2016}. 
This is usually related to the design of the wind turbine structure which is subjected to severe dynamic loading over the waters~\cite{Liu-2016}. 
In this respect, FOWTs tend to have larger rotor diameters and a more slender tower than its land-based counterpart, thus leading to an increase of dynamic loading on the turbine.  
Moreover, FOWTs are usually erected in deep waters with limited accessibility, which potentially exacerbate repair and maintenance issues.

In general, the majority of the loads on wind turbine rotors, showing a periodic nature, is caused by wind shear, tower shadow, turbulence and gravity~\cite{liu2020faulttolerant}.
Individual Pitch Control (IPC) is an effective and widely-used approach to minimize these periodic loads~\cite{Barlas_2010}.
In practice, the pitch control system is subjected to the constraints which are imposed by the physical restrictions of the pitch actuator~\cite{Vali_2016}.
The pitch actuator can only operate within its physical limits while the pitch rate should never exceed the ultimate limitations for safety reasons.
A high pitch rate will increase the cyclic fatigue loading on the pitch actuator and consequently shorten its lifespan. 
In addition, other safety limitations, environmental regulations and wind farm manufacturer specifications may impose additional constraints on the pitch control system.
Exceeding these constraints may result in damage to the pitch actuators and ultimately in the failure of the entire pitch control system. 

This paper introduces a method called constrained Subspace Predictive Repetitive Control (SPRC) to include the physical constraints into a pitch controller for load reduction. 
The basic idea of SPRC was initially proposed by van Wingerden et al.~\cite{Wingerden_2011}, which showed promising results in numerical study~\cite{Navalkar_2014} and in wind tunnel experiments~\cite{Navalkar_2015, frederik2018}.
Since it is a fully data-driven approach consisting of subspace identification~\cite{vanderVeen_2013} and repetitive control, it was also extended for adaptive fault-tolerant control~\cite{liu2020faulttolerant, liu_2020_adaptive, liu2020fast}.
The subspace identification, based on an online solution, is used to derive a linear approximation of wind turbine dynamics.
Then, a predictive repetitive control law is formulated to reduce the dynamic loads under varying operating conditions.
In order to explicitly incorporate the pitch actuator in the control problem formulation, in this paper a Model Predictive Control (MPC) paradigm~\cite{qin_2003} is combined with the SPRC algorithm, so that the future constraint violations can be anticipated and prevented.
By including the repetitive control objectives in a cost function, this technique is able to adapt the control law to specified pitch actuator constraints, actual varying operating conditions and given objectives simultaneously. 
The capability of dealing with the safety constraints is then demonstrated via case studies.
The main attraction of the constrained SPRC is the capability to respect physical and safety restrictions of the pitch control system in FOWTs during operating conditions without compromising the control performance.

The structure of this paper is organized as follows. 
Section \ref{sec:2} introduces the wind turbine model and the simulation environment. 
In Section \ref{sec:3}, the methodology of the constrained SPRC is elaborated. Then, case studies are carried out to verify the developed constrained SPRC in Section \ref{sec:4}. Conclusions are drawn from these case studies and discussed in Section \ref{sec:5}.

\section{Wind turbine model}\label{sec:2}

In this section, the wind turbine model used for the case studies is briefly introduced. 
It is based on the DTU 10MW three-bladed variable speed reference wind turbine integrated with the TripleSpar floating platform~\cite{Bak_2013,Lemmer_2016}.

%

\begin{figure}
\centering 
\includegraphics[width=0.9\columnwidth]{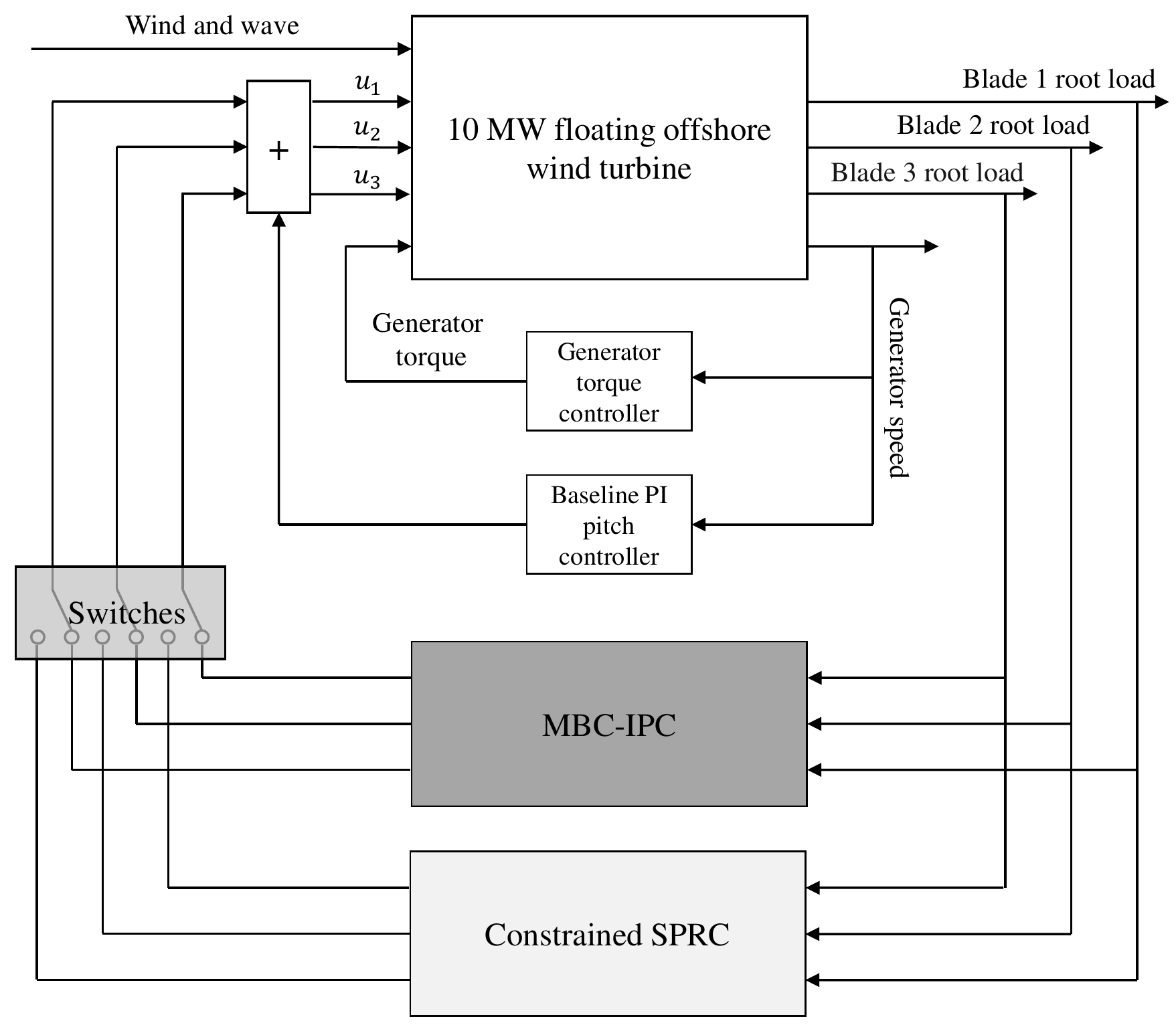}
\caption{Block diagram of the controller and loop of the 10MW FOWT model. 
Three pitch control strategies, \emph{i.e.}, baseline controller~\cite{Jonkman-2005}, MBC-based IPC \cite{Mulders_2019} and constrained SPRC developed in the present study are included for comparisons.}
\label{Pic_block} %
\end{figure}

%
Based on this wind turbine model, the realization of the case studies is illustrated in Fig.~\ref{Pic_block}.
The aero-hydro-structural dynamic part of the wind turbine is simulated via the NREL's Fatigue, Aerodynamics, Structures, and Turbulence (FAST) model~\cite{Jonkman-2005}.
On the other hand, three pitch control strategies are utilized for comparisons, including the baseline controller~\cite{Jonkman-2005}, Multi-Blade Coordinate (MBC)-based IPC~\cite{Mulders_2019} and the proposed  constrained SPRC.
The overall structure of the FOWT model with controllers are depicted in Fig.~\ref{Pic_block}.



Note that the baseline controller is always activated to formulate the collective pitch angles. 
In the IPC cases, one of the IPC strategies is switched on, where the IPC actions are superimposed on top of the collective pitch angles. 
The proposed constrained SPRC, which will be introduced in Section \ref{sec:3}, is carried out to show the significant levels of performance improvement with respect to the other two pitch controllers.

\section{Constrained subspace predictive repetitive control}\label{sec:3}
\begin{figure}[b]
\centering 
\includegraphics[width=1\columnwidth]{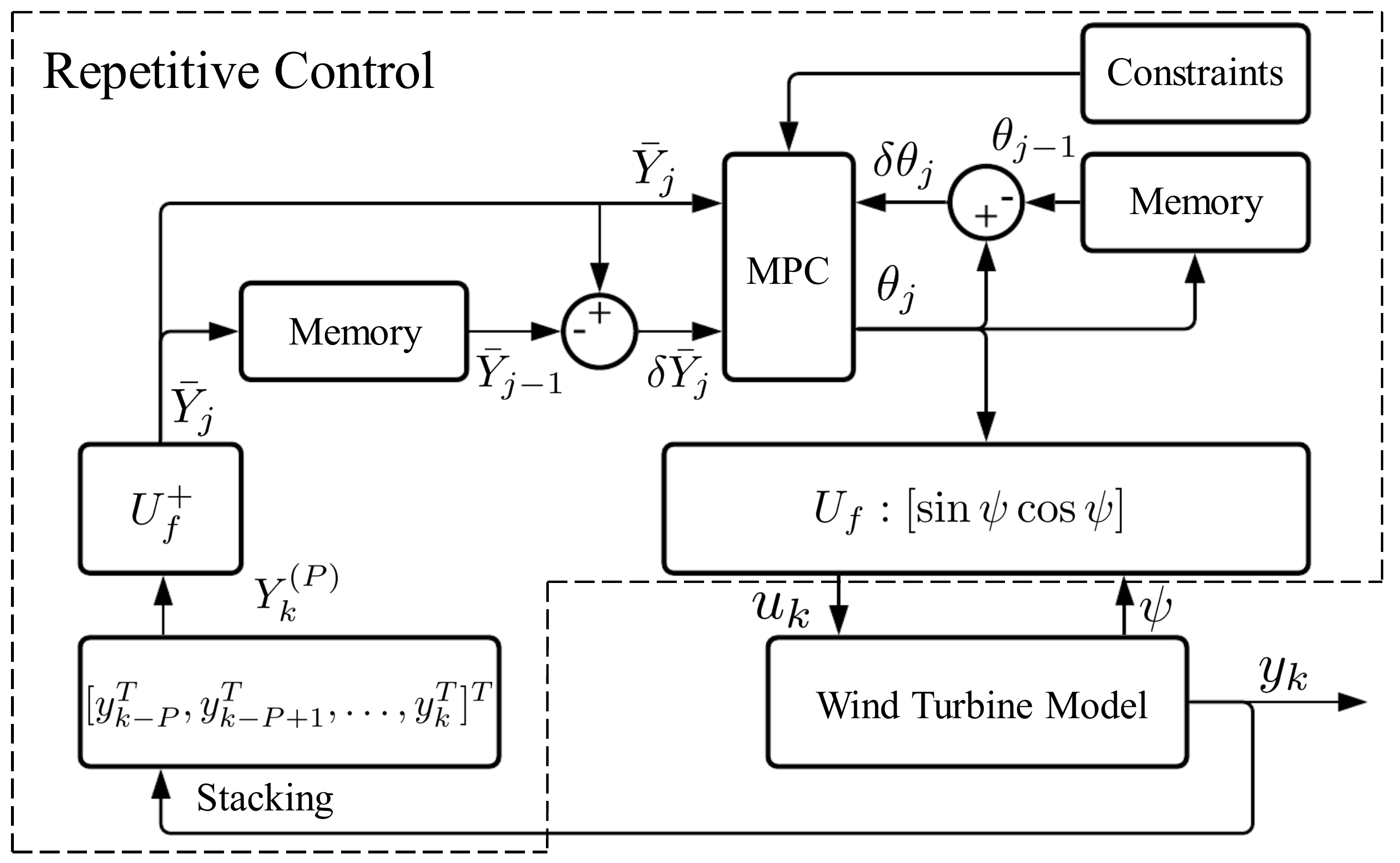}
\caption{Implementation of the repetitive control in SPRC, in which MPC optimization is used to formulate the constrained control law.}
\label{Fig_RC} %
\end{figure}
This section outlines the procedure for formulating the constrained SPRC law, including online recursive subspace identification and receding horizon repetitive control.

\subsection{Online recursive subspace identification}
In the constrained SPRC framework, the wind turbine dynamics is represented by a LTI system affected by unknown periodic disturbances~\cite{Houtzager_2013}.
In prediction form, it can be formulated as 
\begin{equation}
\begin{cases}
\label{eq:predictor:f}
x_{k+1} \!\! &= \tilde{A}x_k+Bu_k+\tilde{E}d_k+Ly_k \\
y_k \!\! &= Cx_k+Fd_k+e_k
\end{cases} \, ,
\end{equation}
where $x_k\in\mathbb{R}^n$, $u_k\in\mathbb{R}^r$ and $y_k\in\mathbb{R}^l$ denote the state, control input and output vectors. 
The number of inputs and outputs are $r = l = 3$.
$u_k$ and $y_k$ represent the blades pitch angles and the blade loads, \emph{i.e.}, the Out-of-Plane bending moment (MOoP) at the blade root at discrete time index $k$. 
Moreover, $d_k\in\mathbb{R}^m$ is the periodic disturbance component of the loads at the blade root, while $e_k\in\mathbb{R}^l$ is the zero-mean white innovation process or the aperiodic component of the blade loads.
$\tilde{A}\triangleq{A-LC}$ and $\tilde{E}\triangleq{E-LF}$, in which $A\in\mathbb{R}^{n\times n}$, $C\in\mathbb{R}^{l\times n}$, $L\in\mathbb{R}^{n\times l}$, $E\in\mathbb{R}^{n\times m}$ and $F\in\mathbb{R}^{l\times m}$ denote the state transition, output, observer, periodic noise input and periodic noise direct feed-through matrices, respectively. $B\in\mathbb{R}^{n\times r}$ represents the input matrix.
Based on the definition of a periodic difference operator $\delta$, the effect of periodic blade loads $d_k$ on the input-output system will be eliminated as
\begin{align*}
 \delta{d}_k &=d_k-d_{k-P}=0 \, ,
\end{align*} 
where $P$ is the period of the disturbance, equalling to the blade rotation period.
Similarly, $\delta{u}$, $\delta{y}$ and $\delta{e}$ can be defined as well.
Applying the $\delta$-notation to \eqref{eq:predictor:f}, this equation can be rewritten as
\begin{equation}
\begin{cases}
\delta{x}_{k+1} \!\! &= \tilde{A}\delta{x}_k+B\delta{u}_k+L\delta{y}_k \\
\delta{y}_k \!\! &= C\delta{x}_k+\delta{e}_k \, 
\end{cases} 
\label{eq:predictor2} \, ,
\end{equation}
where the periodic blade load term disappears.
Then, a stacked vector $\delta{U}^{(p)}_{k}$ for a past time window with the length of $p$ is defined as
\begin{equation}
\delta{U}^{(p)}_{k}=
\left[ \begin{array}{c}
 u_k-u_{k-P}\\
u_{k+1}-u_{k-P+1} \\
\vdots \\
u_{k+p-1}-u_{k+p-P-1}
\end{array} 
\right ]\, .
\label{eq:stacked u}
\end{equation}
Similarly, the vector $\delta{Y}^{(p)}_{k}$ is defined. 
Note $p$ need to be selected large enough, such that $\tilde{A}^{j}\approx0$ $\forall{j}\geq{p}$~\cite{Chiuso_2007}. 
Based on $\delta{U}^{(p)}_{k}$ and $\delta{Y}^{(p)}_{k}$, the future state vector $\delta{x}_{k+p}$ can be approximated as
\begin{equation}
\delta{x}_{k+p} \approx
\left[ \begin{array}{cc}
K^{(p)}_u & K^{(p)}_y 
\end{array} 
\right ]
\left[ \begin{array}{c}
\delta{U}^{(p)}_{k} \\
\delta{Y}^{(p)}_{k} \\
\end{array} 
\right ] \, ,
\label{eq:lifted2}
\end{equation}
in which $K^{(p)}_u$ and $K^{(p)}_y$ are
\begin{align*}
&\ K^{(p)}_u=
\left[ \begin{array}{cccc}
\tilde{A}^{p-1}B & \tilde{A}^{p-2}B & \cdots & B 
\end{array} 
\right ]\,,\\ 
&\ K^{(p)}_y=
\left[ \begin{array}{cccc}
\tilde{A}^{p-1}L & \tilde{A}^{p-2}L & \cdots & L
\end{array} 
\right ]\, .
\label{eq:K}
\end{align*}
By substituting \eqref{eq:lifted2} into \eqref{eq:predictor2}, the approximation of $\delta{y}_{k+p}$ is derived as
\begin{equation}
\delta{y}_{k+p} \approx
\underbrace{
\left[ \begin{array}{cc}
CK^{(p)}_u & CK^{(p)}_y 
\end{array} 
\right ]}_{\Xi}
\left[ \begin{array}{c}
\delta{U}^{(p)}_{k} \\
\delta{Y}^{(p)}_{k} \\
\end{array} 
\right ]
+\delta{e_{k+p}} \, .
\label{eq:lifted3}
\end{equation}
From \eqref{eq:lifted3}, it can be seen that the Markov matrix $\Xi$, includes all the necessary information on the behaviour of the wind turbine model. 
It is determined by the input vector $u^{(r)}$ and output vector $y^{(l)}$. 
In essence, the subspace identification aims to find an online solution of the following Recursive Least-Squares (RLS) optimization problem~\cite{vanderVeen_2013}
\begin{equation}
\hat{\Xi}_k=\text{arg}\min_{\hat{\Xi}_k}\sum_{i=-\infty}^{k}\left \| \delta{y}_i-\lambda\hat{\Xi}_k
\left[ \begin{array}{c}
\delta{U}^{(p)}_{i-p} \\
\delta{Y}^{(p)}_{i-p} \\
\end{array} 
\right ]
\right \| ^2_2 \, .
\label{eq:Markov parameters2}
\end{equation}
In \eqref{eq:Markov parameters2}, $\lambda$ is a forgetting factor ($0\ll\lambda\leq{1}$) to alleviate the effect of past data, and adapt to the varying system dynamics online.
In this paper, a value close to 1, \emph{i.e.}, $\lambda=0.99999$, is chosen for the optimization process.
To obtain a unique solution to this RLS optimization problem, a persistently exciting signal is superimposed on the top of the wind turbine control input.
Subsequently, the RLS optimization \eqref{eq:Markov parameters2} is implemented with a QR algorithm~\cite{Sayed_1998} in an online recursive manner to obtain $\hat{\Xi}_{k}$.
The estimates of $\hat{\Xi}_k$ are then used in an MPC algorithm to formulate a repetitive control law subjected to the actual pitch actuator constraints.
The implementation of the repetitive control problem formulation will be elaborated in the next subsection.

\subsection{Receding horizon repetitive control}
In the repetitive control, the control law is predicted over $P$. Considering that $P\geq{p}$ and usually $P$ is much larger than $p$, the output equation can be lifted over $P$ as
\begin{multline}
{Y}^{(P)}_{k+P}=
\left[ \begin{array}{ccc}
I_{l\cdot{P}} & \Gamma^{(P)} \widehat{K^{(P)}_u} & \Gamma^{(P)} \widehat{K^{(P)}_y} \\
\end{array} 
\right ]
\left[ \begin{array}{c}
Y^{(P)}_k \\
\delta{U}^{(P)}_k \\
\delta{Y}^{(P)}_k 
\end{array} 
\right ]
\\
+\hat{H}^{(P)}\delta{U}^{(P)}_{k+P} \, ,
\label{eq:lift:y3}
\end{multline}
where it holds $\Gamma^{(P)}=(I-\tilde{G}^{(P)})^{-1}\tilde{\Gamma}^{(P)}$ and $\hat{H}^{(P)}=(I-\tilde{G}^{(P)})^{-1}\tilde{H}^{(P)}$.
The Toeplitz matrix $\tilde{H}^{(P)}$ is defined as
\begin{equation*}
\tilde{H}^{(P)}=
\left[ \begin{array}{cccc}
0 & 0 & 0 & \cdots  \\
CB & 0 & 0 & \cdots  \\
C\tilde{A}B & CB & 0 & \cdots \\
\vdots & \vdots & \ddots & \vdots \\
C\tilde{A}^{p-1}B & C\tilde{A}^{p-2}B & C\tilde{A}^{p-3}B & \cdots \\
0 & C\tilde{A}^{p-1}B & C\tilde{A}^{p-2}B & \cdots \\
0 & 0 & C\tilde{A}^{p-1}B & \cdots \\
\vdots & \vdots & \ddots & \ddots \\
\end{array} 
\right ]
\, .
\label{eq:H}
\end{equation*}
By replacing $B$ with $L$, $\tilde{G}^{(P)}$ can be defined as well.
The extended observability matrix $\tilde{\Gamma}^{(P)}$ is defined as
\begin{equation*}
\tilde{\Gamma}^{(P)}=
\left[ \begin{array}{c}
C \\
C\tilde{A} \\
C\tilde{A}^2 \\
\vdots \\
C\tilde{A}^p \\
0 \\
\vdots \\
0
\end{array} 
\right ]
\, .
\label{eq:Gamma}
\end{equation*}
Next, a basis function projection~\cite{Wingerden_2011} is employed to limit the spectral content of the pitch control input within the frequency range of interest, which will also reduce the dimension of \eqref{eq:lift:y3} that must be solved in the MPC framework.
The transformation matrix of the basis function projection for the 1P frequency is defined as
\begin{equation}
\phi=
\underbrace{
\left[ \begin{array}{cccc}
\sin{\psi} & \cos{\psi} 
\end{array} 
\right ]}_{U_f} 
\otimes{I}_r
\, ,
\label{eq:with basis function}
\end{equation}
where $U_f$ denotes the basis function while $\psi$ is the rotor azimuth. 
The mathematical symbol $\otimes$ represents the Kronecker product.
Based on the basis function, the control inputs at specific frequencies can be synthesised as
\begin{equation}
U_j=\phi \cdot \theta_j
\, ,
\label{eq:control input}
\end{equation}
where $j=0,1,2,\cdots$ represents the rotation count. $\theta\in\mathbb{R}^{b r}$, which determines the amplitudes and phase of the sinusoids, is updated at each $P$, while $b$ corresponds to the column number of $\phi$.
Similarly, the output can be transformed onto the subspace that defined by the basis function, as
\begin{equation}
\bar{Y}_j=\phi^{+}Y^{(P)}_j
\, ,
\label{eq:control output}
\end{equation}
in which the symbol $+$ is the Moore-Penrose pseudo-inverse. 
Based on the basis function projection, \eqref{eq:lift:y3} is subsequently reformulated into a lower dimensional state-space model as
\begin{multline}
\underbrace{
\left[ \begin{array}{c}
\bar{Y}_{j+1}\\
\delta{\theta}_{j+1} \\
\delta{\bar{Y}}_{j+1} \\
\end{array} 
\right ]}_{\bar{\mathcal{K}}_{j+1}}
=
\underbrace{
\left[ \begin{array}{ccc}
I_{l\cdot{b}} & \phi^{+}\Gamma^{(P)} \widehat{K^{(P)}_u}\phi & \phi^{+}\Gamma^{(P)} \widehat{K^{(P)}_y}\phi \\
0_{l\cdot{b}} & 0_{r\cdot{b}} & 0_{l\cdot{b}} \\
0_{l\cdot{b}} & \phi^{+}\Gamma^{(P)} \widehat{K^{(P)}_u}\phi & \phi^{+} \Gamma^{(P)}\widehat{K^{(P)}_y}\phi 
\end{array} 
\right]}_{\bar{A}_j}
\\
\underbrace{
\left[ \begin{array}{c}
\bar{Y}_j \\
\delta{\theta}_j \\
\delta{Y}_j
\end{array} 
\right]}_{\bar{\mathcal{K}}_j}
+
\underbrace{
\left[ \begin{array}{c}
\phi^{+}\hat{H}^{(P)}\phi \\
I_{r\cdot{b}} \\
\phi^{+}\hat{H}^{(P)}\phi
\end{array} 
\right]}_{\hat{B}_j}
\delta{\theta}_{j+1}
\, .
\label{eq:state-space form_lower}
\end{multline}
Following the philosophy of the MPC algorithm, the control objectives can be introduced in the cost function as
\begin{multline}
J(\bar{\mathcal{K}},\mathbf{U})=\sum_{i=0}^{N_p} 
(\bar{\mathcal{K}}_{j+i|j})^{T}Q\bar{\mathcal{K}}_{j+i|j}+ 
\\
\sum_{i=0}^{N_u - 1} (\delta\theta_{j+i|j})^{T}R \delta\theta_{j+i|j}
\, ,
\label{eq:cost function}
\end{multline}
while the goal function being optimizer for is
\begin{equation}
V(\bar{K_j}) = \min_{\mathbf{U}} J(\bar{\mathcal{K}}_{j},\mathbf{U})
\, ,
\label{eq:MPC optimization problem}
\end{equation}
where $Q$ and $R$ denote positive-definite weighting matrices, while $N_p$ and $N_u$ are the prediction and control horizons, respectively.
$\mathbf{U} \triangleq [\delta\theta^T_{j}, \cdots, \delta\theta^T_{j+N_u-1} ]\in \mathbb{R}^{b r \times N_u}$ is a sequence of future control actions. They are computed by the MPC optimization process over the prediction horizon at each $j$. 
This will optimize the future behavior of the wind turbine while respecting the pitch actuator constraints.

With the definition of $\delta \theta_{j+1} = \theta_{j+1} - \theta_{j}$, the following pitch actuator constraints are imposed on \eqref{eq:state-space form_lower}
\begin{equation}
\begin{cases}

- \bar{U}_j - \delta U_{\text{max}} \cdot \Delta T
\leq
\phi \cdot \delta \theta_{j+1} \leq

\delta U_{\text{max}} \cdot \Delta T - \bar{U}_j \\

- \bar{U}_j - \phi \cdot \theta_j
\leq 
\phi \cdot \delta \theta_{j+1} \leq

U_{\text{max}} - \bar{U}_j - \phi \cdot \theta_j
\end{cases}
\, ,
\label{eq:MPC constraints_reformulation}
\end{equation}
where $\bar U_k$ denotes the collective pitch angles from the baseline controller. $\Delta T$ is the fixed discrete time step. $\delta U_{\text{max}}$ and $U_{\text{max}}$ represent the limitations of pitch rate and angle, respectively. Both of them are usually determined by wind farm manufacturer specifications, safety limitations and environmental regulations. 
From \eqref{eq:MPC constraints_reformulation}, it can be seen that the constraints of the future $\delta \theta_{j+1}$ in the prediction horizon are actually determined by $\theta_{j}$ and $\bar{U}_j$ at last rotation count $j$.

Only the first element $\delta \theta_j^T$ in the vector of the optimal input consequence $\mathbf{U}$ is actually selected while the remaining elements are discarded. 
In this paper, $U_{\text{max}}$ and $\delta U_{\text{max}}$ will be used as a design parameter to demonstrate the constraints handling capability of the proposed constrained SPRC methodology.
Considering that $\delta \theta_{j+1} = \theta_{j+1} - \theta_{j}$, $\theta_j^T$ can be synthesised. 
As a consequence, the repetitive control signals $U_{k+1}$ at time step $k+1$ are finally computed according to \eqref{eq:control input}.
At the next rotation count, the state $\bar{\mathcal{K}}_{j+1}$ will be updated and used to be an initial condition of the MPC optimization.
Following the philosophy of the receding horizon principle~\cite{qin_2003}, the cost function in \eqref{eq:cost function} will roll ahead one step and all the procedure will be repeated.

Equation~\eqref{eq:MPC optimization problem} can be solved as a standard Quadratic Programming (QP) problem, by converting the control objectives in \eqref{eq:cost function} in the following form
\begin{equation}
J(\bar{\mathcal{K}}_j,\mathbf{U}) = X^T \mathcal{Q} X + \mathbf{U}^T \mathcal{R} \mathbf{U}
\, ,
\label{eq:QP form}
\end{equation}
where $X = [\bar{\mathcal{K}}_{j}, \bar{\mathcal{K}}_{j+1}, \cdots, \bar{\mathcal{K}}_{j+N_p}]^T$ corresponds to the vector of state predictions. $\mathcal{Q}$ and $\mathcal{R}$ are the weight matrices, which are
\begin{equation}
\mathcal{Q} = \text{diag}(Q, \cdots, Q) \,\,\,\,\,\,  \mathcal{R} = \text{diag}(R, \cdots, R)
\, .
\label{eq:QR}
\end{equation}

By introducing the following prediction matrices
\begin{align}
\mathcal{A}
= 
\left[ \begin{array}{c}
I \\
\bar{A}_j \\
\vdots  \\
{\bar{A}_j}^{N_u} \\
\vdots  \\
{\bar{A}_j}^{N_p}
\end{array} \right]
\,,
\mathcal{B} = 
\left[ \begin{array}{ccc}
0 & \cdots & 0 \\
\hat{B}_j & \cdots & 0 \\
\vdots & \ddots & \vdots \\
{\bar{A}_j}^{N_u-1}\hat{B}_j & \cdots & \hat{B}_j \\
\vdots & \vdots & \vdots \\
{\bar{A}_j}^{N_p-1}\hat{B}_j & \cdots & \sum_{i=0}^{N_p - N_u} {\bar{A}_j}^i \hat{B}_j
\end{array} \right]
\, ,
\end{align}
the following predictive system is derived, that is
\begin{equation}
X = \mathcal{A}\bar{\mathcal{K}}_j + \mathcal{B} \mathbf{U}
\, .
\label{eq:predictive system}
\end{equation}

Combining \eqref{eq:predictive system} with \eqref{eq:cost function}, the MPC optimization is solved by the following QP problem
\begin{multline}
V(\bar{\mathcal{K}}_j) = \bar{\mathcal{K}}_j^T \mathcal{Y} \bar{\mathcal{K}}_j + \min_{\mathbf{U}} \{\mathbf{U}^T H \mathbf{U} + 2\bar{\mathcal{K}}_j^T F \mathbf{U}\}
\, \,\,\,\,\,\,
\\
\text{subject to} \qquad \qquad \qquad \qquad \qquad \qquad \,\,
G \mathbf{U} \leq W
\, ,
\label{eq:QP form2}
\end{multline}
where $H=\mathcal{B}^T \mathcal{Q} \mathcal{B} + \mathcal{R}$, $F = \mathcal{A}^T \mathcal{Q} \mathcal{B}$ and $\mathcal{Y} = \mathcal{A}^T \mathcal{Q} \mathcal{A}$.
In addition, $G$ and $W$ are defined according to \eqref{eq:MPC constraints_reformulation}, in a similar manner as in the paper~\cite{Bemporad_2002}.

The block diagram of the repetitive control is schematically presented in Fig. \ref{Fig_RC}. In summary, the Markov matrix $\hat{\Xi}_{k}$ is obtained by solving the RLS optimization in an online recursive manner \eqref{eq:Markov parameters2}.
Then, the constrained repetitive control law is formulated in \eqref{eq:control input} by solving an MPC optimization problem \eqref{eq:MPC optimization problem}-\eqref{eq:QP form2}.



\section{Case study}\label{sec:4}
In this section, the effectiveness of the developed constrained SPRC is demonstrated on the 10MW FOWT model via a series of case studies. 

\subsection{Model configuration}
The wind turbine model, which has been introduced in Section \ref{sec:2}, is simulated by the \emph{FAST v8.16} tool~\cite{Jonkman-2005}.
Both laminar and turbulent wind conditions are considered, as shown in Table~\ref{table:load_case}.
The turbulent varying wind field is simulated with the TurbSim model\footnote{TurbSim: a stochastic inflow turbulence tool to generate realistic turbulent wind fields. https://nwtc.nrel.gov/TurbSim.\label{TurbSim}}, where the Turbulence Intensity (TI) is specified to be 3.75\%. 
In total, eight Load Cases (LCs) are considered in the simulations. 

\begin{table}[h]
\setlength{\tabcolsep}{2.9mm}
{
\caption{Environmental conditions and predefined pitch actuator constraints (\emph{i.e.}, maximum pitch angle, pitch rate) in all LCs.\label{table:load_case}}
\vspace{-0.3cm}
\begin{tabular}{cccc}
\hline \hline
\textbf{Load case} & \textbf{Wind speed} [m/s] & $U_{\text{max}}$ [deg] & $\delta U_{\text{max}}$ [deg/s] \\ \hline
$1$  &  12    & 4.9   & 1.1 \\ 
$2$  &  12    & 4.8   & 0.5    \\
$3$  &  16    & 12.9  & 1.0 \\ 
$4$  &  16    & 13.1  & 0.2    \\
$5$  &  20    & 18.4  & 0.9 \\ 
$6$  &  20    & 18.1  & 1.5    \\
$7$  &  16 (TI: 3.75\%)   & 13.5  & 8.0   \\
$8$  &  16 (TI: 3.75\%)   & 20.0    & 1.5 \\ \hline \hline
\end{tabular}
}
\end{table}
\subsection{Results and discussions}
\begin{figure}
\centering 
\includegraphics[width=0.85\columnwidth]{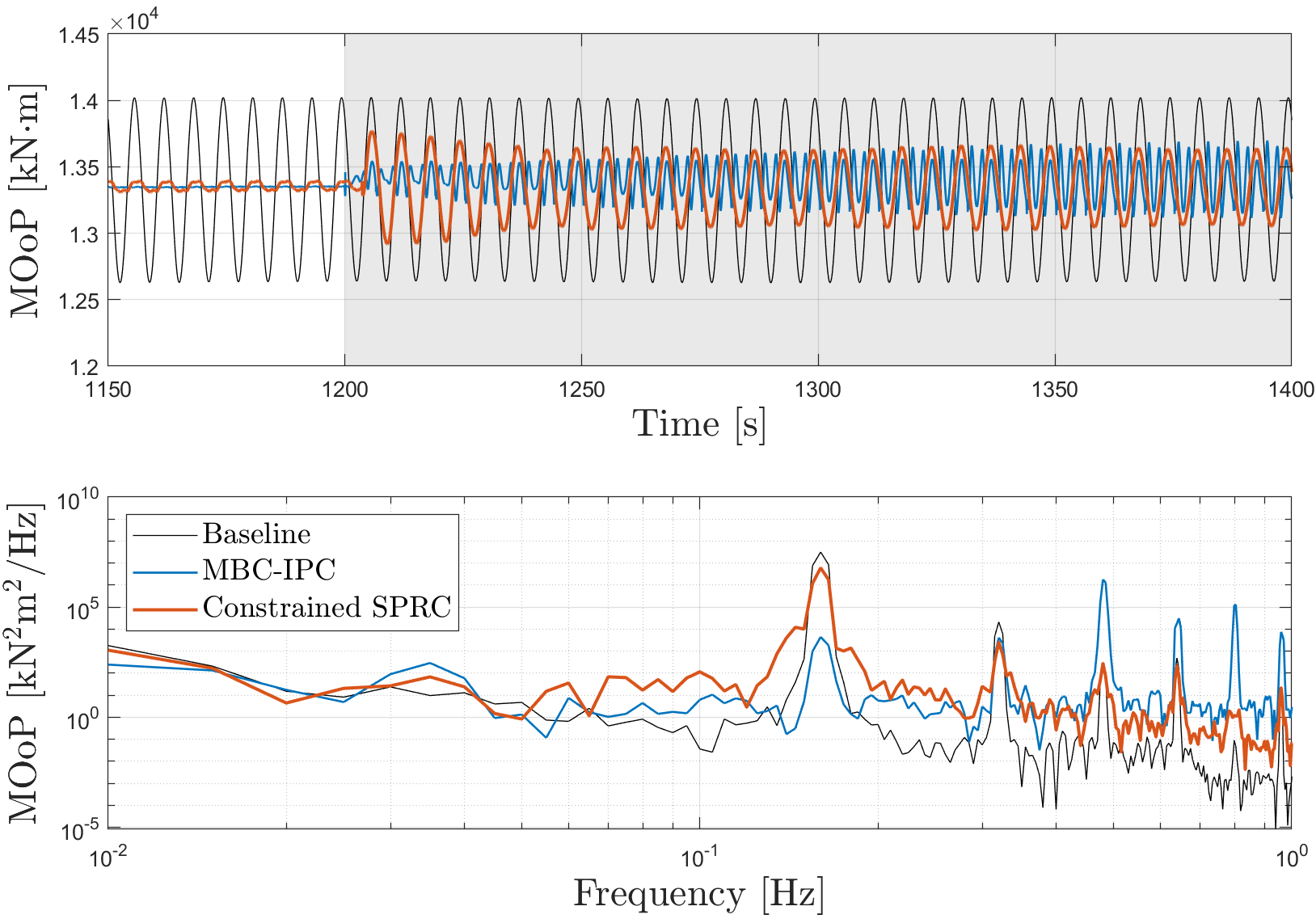}
\caption{MOoP on the blade root in LC3, where the pitch actuator constraints in the developed constrained SPRC are added at 1200s. Note that time periods of the pitch actuator constraints are indicated by a grey background. Blades 2-3 are not shown since they present similar results as blade 1. Both will be not repeated in other figures.}
\label{Fig_MOoP} %
\end{figure}

\begin{figure}
\centering 
\includegraphics[width=0.85\columnwidth]{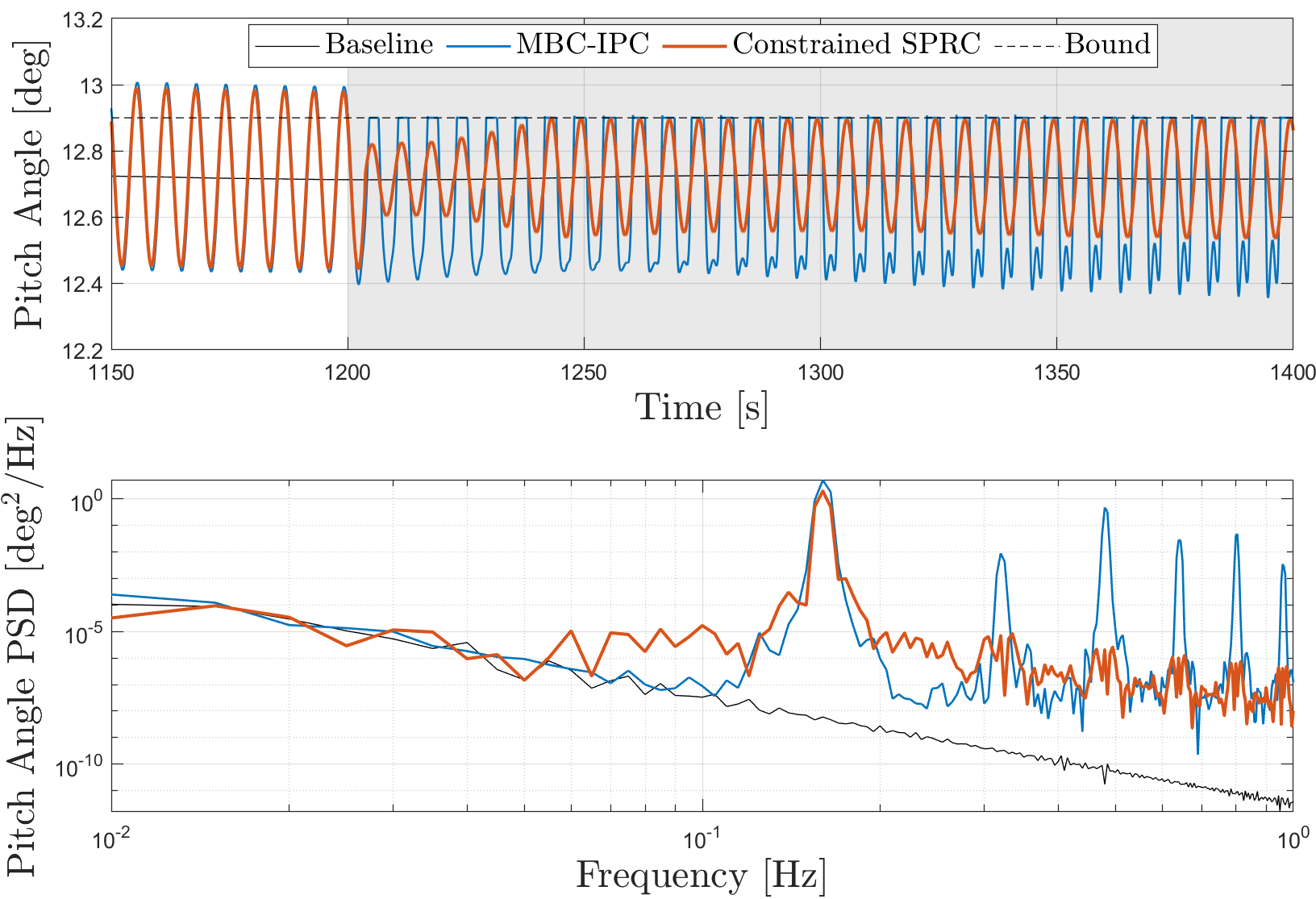}
\caption{Pitch angle in LC3, where the pitch actuator constraints in the developed constrained SPRC are added at 1200s.}
\label{Fig_pitch_angle} %
\end{figure}

\begin{figure}
\centering 
\includegraphics[width=0.85\columnwidth]{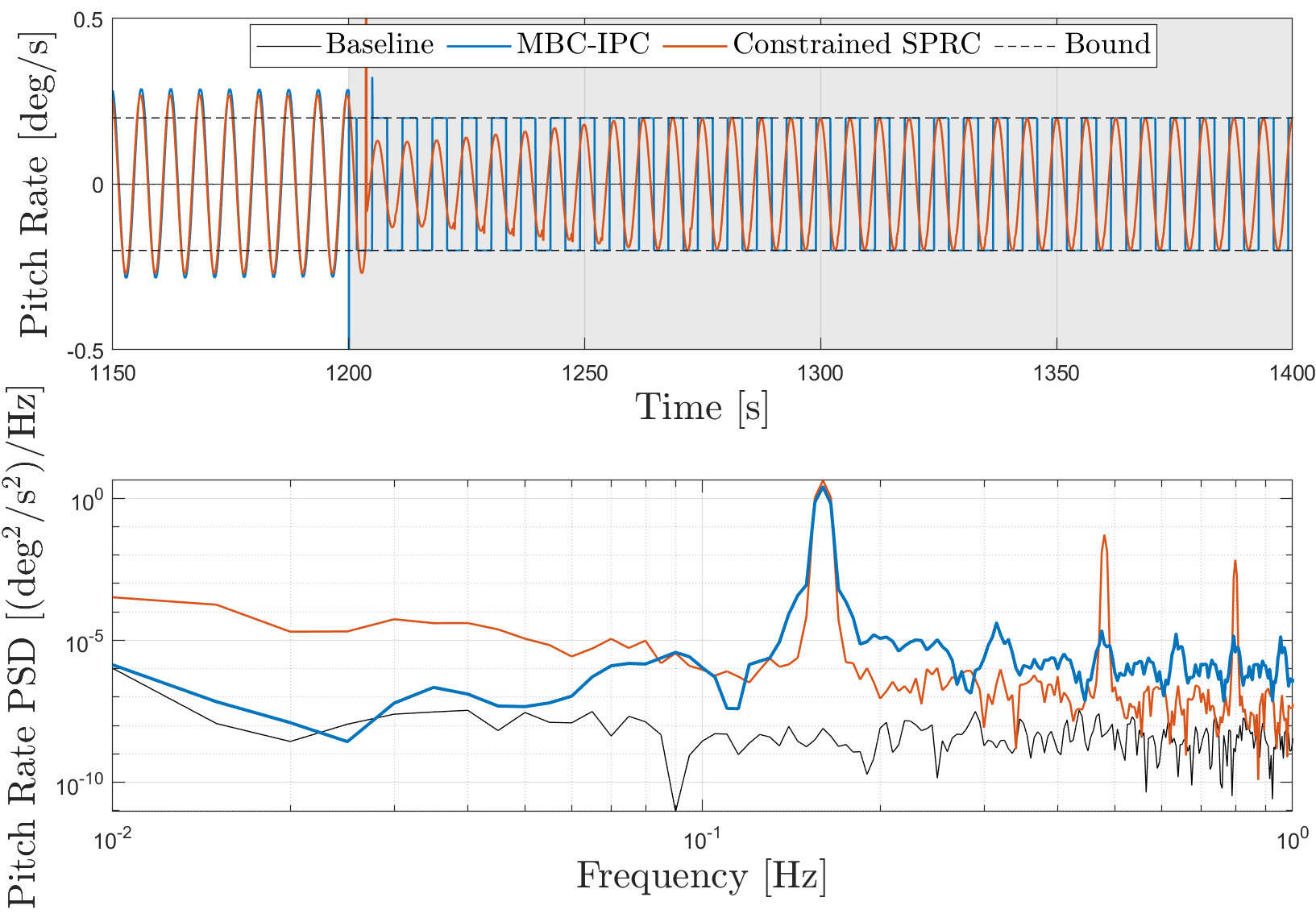}
\caption{Pitch rate in LC4, where the pitch actuator constraints in the developed constrained SPRC are added at 1200s.
\textbf{Remark 1.} The spike of the signals at 1200s is induced by the transition between the simulations of constraints and non-constraints.}
\label{Fig_pitch_rate} %
\end{figure}

\begin{figure}
\centering 
\includegraphics[width=0.85\columnwidth]{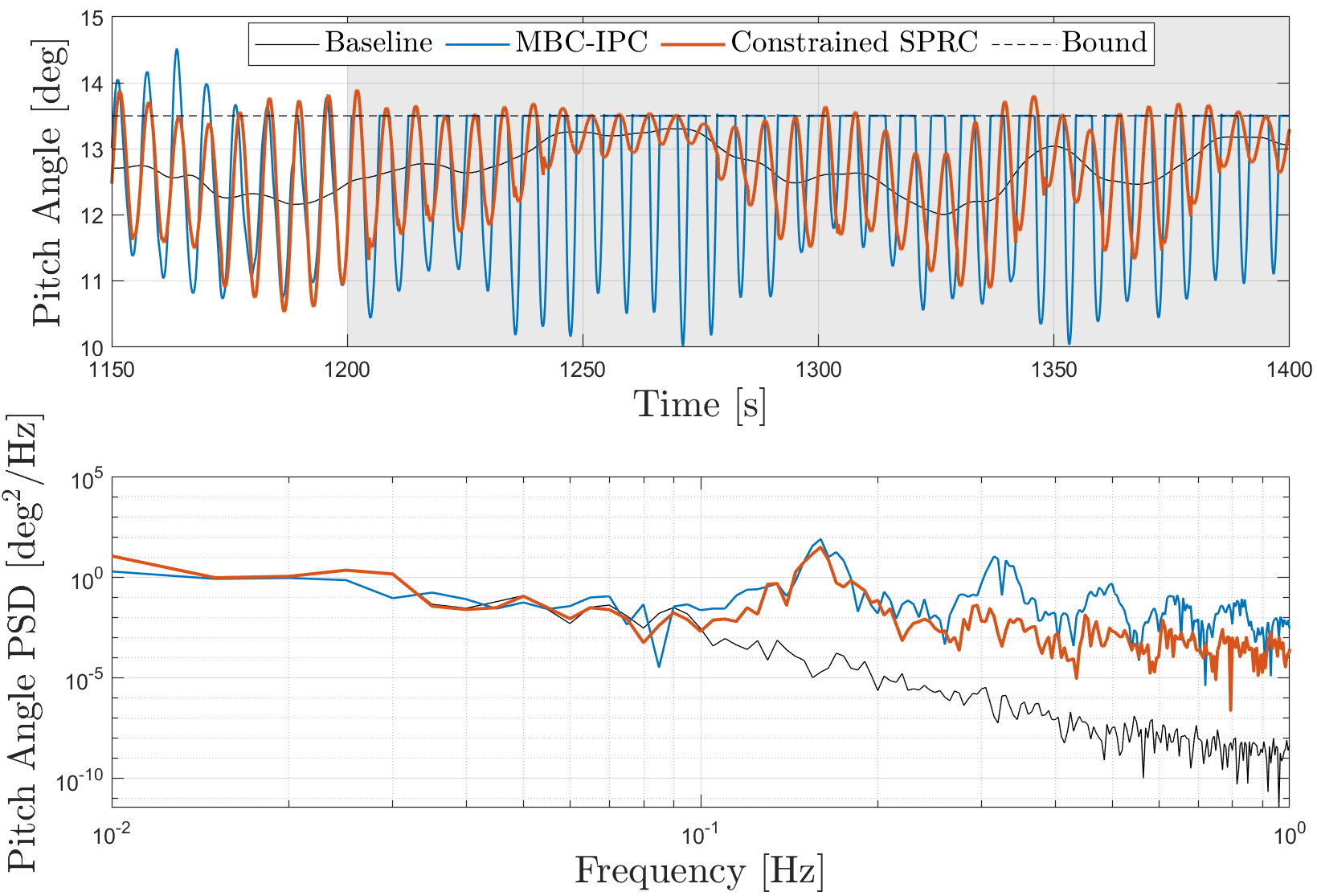}
\caption{Pitch angle in LC7, where the pitch actuator constraints in the developed constrained SPRC are added at 1200s.}
\label{Fig_pitch_angle_tur} %
\end{figure}

Figs. \ref{Fig_MOoP}-\ref{Fig_pitch_angle_tur} show the SPRC capability dealing with constraints in LCs 3-4, 7.
It is observed that in the proposed constrained SPRC, all the input constraints are successfully incorporated into the control problem formulation.
In the turbulent case, the pitch angles are slightly over the bound at around 1350s, as presented in Fig.~\ref{Fig_pitch_angle_tur}. This is actually caused by the loss of information in the basis function projection. 
However, the MBC-based IPC, on the basis of the Coleman transformation, is not able to take into account the limitations of the pitch actuator. Instead, a saturation block is used to clip the the pitch angles at the maximum/minimum allowable value, as illustrated in Fig.~\ref{Fig_pitch_angle}.
Such an implementation is not preferred in practice since it will introduce high load harmonics.
For example, the Power Spectral Densities (PSDs) of pitch angles derived from MBC-based IPC in Fig.~\ref{Fig_pitch_angle} are around $\sim 0.47  \text{deg}^2/$Hz at 0.48\,Hz, which is much larger than $\sim 1\times 10^{-10} \text{deg}^2/$Hz in the baseline controller case.
When the constrained SPRC is used, the shape of the pitch angles is preserved while the amplitudes are reduced to $\sim 6.3\times 10^{-7} \text{deg}^2/$Hz, and thus prevent violating constraints. 
Therefore, the proposed constrained SPRC is effective to avoid violating constraints while the sinusoidal shape of the pitch angles can be successfully maintained.
The conventional IPC using the saturation may introduce high harmonics by simply cutting off the exceeded pitch angles. 

Similar results are observed in the other LCs.
For the sake of comparisons, all the results are summarized in Table~\ref{table:ADC}.
Specifically, a criterion called Actuator Duty Cycle (ADC)~\cite{Bottasso_2013}, which is widely used to estimate the lifespan of pitch actuators, is utilized to quantify the pitch activities.
Table~\ref{table:ADC} illustrates the pitch ADC for all the simulation runs.
\begin{table}
\setlength{\tabcolsep}{3.8mm}
{
\caption{Comparisons of ADC with different control strategies*. \label{table:ADC}}
\vspace{-0.3cm}
\begin{tabular}{lcccc}
\hline  \hline
\textbf{ADC} [\%]
& $\textbf{LC1}$   & $\textbf{LC2}$   & $\textbf{LC3}$  &$\textbf{LC4}$   
\\ 
Constrained SPRC   & 12.86    & 17.07  & 10.80    &  60.39 \\
MBC-IPC          & 16.39     & 56.74  & 18.12   &  103.21 \\ 
\hline 
& $\textbf{LC5}$  & $\textbf{LC6}$ & $\textbf{LC7}$  &   $\textbf{LC8}$ \\ 
Constrained SPRC   & 26.03    & 4.62  & 5.99    &  55.36 \\
MBC-IPC          & 31.40     & 35.14  & 11.65   &  62.85 \\ 
\hline \hline
\end{tabular}}

\footnotesize
\vspace{0.1cm}
*These results are calculated based on the data $1200\,\text{s}- 1400\,\text{s}$.
\end{table}
It is evident that the proposed method, in general, leads to lower pitch ADC on all the blade pitch actuators than MBC-based IPC.
For instance, the constrained SPRC only shows $\sim 60\%$ pitch ADC in LC4 where tight constraints are imposed on the actuators. This implies a big decrease of pitch activities compared to the pitch ADC in MBC-based IPC. 
Averaging over all the cases, the pitch ADC is reduced by the constrained SPRC by $\sim42\%$.
The most evident reduction of pitch ADC occurs in the cases with tight constraints such as in LC6, which shows a maximum of $\sim87\%$.

In conclusion, these results exhibit that the proposed constrained SPRC is able to significantly reduce the pitch ADC by incorporating the safety constraints into the control problem formulation.
The conventional MBC-based IPC is not an effective way to respect the constraints, which shows higher pitch ADC caused by the undesired load harmonics at high frequencies. 






\section{Conclusions}\label{sec:5}
This paper presents a viable solution to incorporate the constraints into the Individual Pitch Control (IPC).
It is achieved by developing a constrained Subspace Predictive Repetitive Control (SPRC). 
It is a fully data-driven approach, in which the actuator constraints are explicitly considered in a Model Predictive Control (MPC) optimization step. 
Case studies demonstrate that the conventional MBC-based IPC is not an effective way to take into account the constraints. By simply clipping the exceeded pitch angles at the minimal allowable value, higher load harmonics are introduced in the wind turbine. 
On the other hand, the proposed constrained SPRC is able to prevent the violation of the constraints while the sinusoidal shape of the pitch angles can be successfully preserved. 
Averaging over all the cases, the constrained SPRC attains a reduction of the pitch actuator duty cycle by $\sim 42\%$ compared to MBC-based IPC.
Therefore, it can be concluded that  the developed constrained SPRC outperforms other control strategies in case the restrictive space of possible pitch control actions are enforced by the constraints.
This will help to ensure the safety of the pitch control system and potentially reduce the damage rate of the wind turbine.


\bibliographystyle{IEEEtran} 
\bibliography{references}


\end{document}